\documentclass[fleqn,usenatbib]{mnras}

\usepackage{newtxtext,newtxmath}
\usepackage[T1]{fontenc}
\DeclareRobustCommand{\VAN}[3]{#2}
\let\VANthebibliography\thebibliography
\def\thebibliography{\DeclareRobustCommand{\VAN}[3]{##3}\VANthebibliography}

\usepackage{hyperref}
\usepackage{graphicx}	% Including figure files
\usepackage{amsmath}	% Advanced maths commands
\usepackage{bm}
\usepackage{fancyhdr}
\title[DM substructure with optical lensed quasars]{Detecting dark matter substructure with lensed quasars in optical bands}

\author[Liu et al.]{
Jianxiang Liu,$^{1,2}$
Kai Liao,$^{3}$\thanks{E-mail: liaokai@whu.edu.cn}
and Yan Gong$^{1,2}$\thanks{E-mail: gongyan@bao.ac.cn}
\\
$^{1}$National Astronomical Observatories, Chinese Academy of Sciences, Beijing 100101, P. R. China\\
$^{2}$School of Astronomy and Space Sciences, University of Chinese Academy of Sciences, Beijing 100049, P. R. China\\
$^{3}$School of Physics and Technology, Wuhan University, Wuhan 430072, China
}

\date{Accepted XXX. Received YYY; in original form ZZZ}

\pubyear{\the\year{}}

\begin{document}
\label{firstpage}
\pagerange{\pageref{firstpage}--\pageref{lastpage}}
\maketitle

% Abstract of the paper
\begin{abstract}
Flux ratios of multiple images in strong gravitational lensing systems provide a powerful probe of dark matter substructure. 
Optical flux ratios of lensed quasars are typically affected by stellar microlensing, and thus studies of dark matter substructure often rely on emission regions that are sufficiently extended to avoid microlensing effects. 
To expand the accessible wavelength range for studying dark matter substructure through flux ratios and to reduce reliance on specific instruments, we confront the challenges posed by microlensing and propose a method to detect dark matter substructure using optical flux ratios of lensed quasars. 
We select 100 strong lensing systems consisting of 90 doubles and 10 quads to represent the overall population and adopt the Kolmogorov--Smirnov (KS) test as our statistical method. 
By introducing different types of dark matter substructure into these strong lensing systems, we demonstrate that using quads alone provides the strongest constraints on dark matter and that several tens to a few hundred independent flux ratio measurements from quads can be used to study the properties of dark matter substructure and place constraints on dark matter parameters. 
Furthermore, we suggest that the use of multi-band flux ratios can substantially reduce the required number of quads.
Such sample sizes will be readily available from ongoing and upcoming wide-field surveys.

\end{abstract}

\begin{keywords}
gravitational lensing: strong -- gravitational lensing: micro -- cosmology: dark matter -- methods: statistical
\end{keywords}

%%%%%%%%%%%%%%%%%%%%%%%%%%%%%%%%%%%%%%%%%%%%%%%%%%

%%%%%%%%%%%%%%%%% BODY OF PAPER %%%%%%%%%%%%%%%%%%

\section{Introduction}
Dark matter is a hypothetical form of matter that constitutes the majority of the matter content in the Universe \citep{2020A&A...641A...6P}. 
Numerous independent astrophysical observations have provided strong evidence for its existence, including the cosmic microwave background \citep{2020A&A...641A...1P}, galaxy rotation curves \citep{1985ApJ...295..305V}, and gravitational lensing \citep{2024SSRv..220...87S}. 
Despite decades of research into its properties, the fundamental nature of dark matter remains unknown. 
Cold dark matter (CDM), exemplified by weakly interacting massive particles (WIMPs), has been highly successful in reproducing the large-scale structure of the Universe \citep{2018MNRAS.475..624N}. 
However, it faces several challenges on small scales, such as the missing satellite problem, the cusp-core problem, and the too-big-to-fail problem \citep{2017ARA&A..55..343B}. 
To address these discrepancies, alternative dark matter models have been proposed, such as warm dark matter  (WDM) \citep{2019PrPNP.104....1B} and fuzzy dark matter (FDM), the latter consisting of ultra-light bosons with masses on the order of $10^{-22}\,\mathrm{eV}$ \citep{2021ARA&A..59..247H}.
These alternative models are also capable of reproducing the large-scale structure of the Universe, but they exhibit different behaviors from CDM on small scales.
Although some studies suggest that these small-scale issues may arise from poorly understood baryonic processes and uncertainties in galaxy formation modeling rather than flaws in the CDM paradigm itself, the debate remains unresolved \citep{2015MNRAS.454.2981C,2017ARA&A..55..343B}.

Observational probes such as the abundance of Milky Way satellite galaxies, the Lyman-$\alpha$ forest, and the density profiles of dwarf galaxies have been used to constrain the properties of dark matter \citep{2017PhRvL.119c1302I,2021PhRvL.126i1101N,2025PhRvL.134o1001Z}. 
Strong gravitational lensing, as a purely gravitational probe, is highly sensitive to the matter distribution between the observer and the source \citep{2023arXiv230611781V}. 
Therefore, it provides a powerful and complementary method for constraining the nature of dark matter.
Specifically, in a strong gravitational lensing system, the light from a background source, typically a quasar, is deflected by a foreground galaxy (or a group or cluster of galaxies) before reaching the observer, resulting in the formation of multiple images.

These images give rise to several observable quantities, including image positions, flux ratios (since the intrinsic luminosity of the source is unknown), and time delays \citep{2024SSRv..220...87S}. 
To use strong lensing as a probe of dark matter, one must construct a mass model of the foreground lens galaxy. 
However, perturbations due to dark matter substructures within the lens, or line-of-sight structures, can cause discrepancies between the predicted and observed image properties. 
These discrepancies are known as anomalies and can be used to infer the properties of dark matter substructure, thereby shedding light on the fundamental nature of dark matter \citep{2023arXiv230611781V}.

The most commonly studied observable in this context is the flux ratio anomaly, which is particularly sensitive to small-scale mass perturbations because it is related to the second derivative of the lensing potential. 
In addition to dark matter substructures, flux ratio anomalies can also arise from other effects such as microlensing by stars, free-free absorption, and dust extinction.
Microlensing by stars occurs because the size of a quasar’s optical accretion disk is comparable to the Einstein radius of a typical star in the lensing galaxy. 
As a result, light emitted from the accretion disk is deflected by individual stars, creating many closely spaced 'micro-images' with separations on the order of a few microarcseconds. 
These micro-images are too closely spaced to be resolved individually, but they can significantly alter the observed fluxes, thereby contributing to flux ratio anomalies \citep{2024SSRv..220...14V}.
To avoid the impact of microlensing, previous studies have typically refrained from using optical flux ratios to probe dark matter substructure. 
Instead, they have focused on flux ratios measured from emission regions that are sufficiently large to be unaffected by microlensing, while remaining sensitive to perturbations from dark matter substructures. 
These include the narrow-line region, radio emission from the background quasar, and the warm dust region of the quasar \citep{2019A&A...623A..79C,2020MNRAS.491.6077G,2024MNRAS.530.2960N,2024MNRAS.535.1652K}.

In this work, we are inspired by the method proposed in \cite{2023A&A...673A..88A}, where the authors employed the Kolmogorov–Smirnov (KS) test to investigate how many pairs of light curves are required to distinguish between scenarios in which primordial black holes (PBHs) constitute 0 and 100 $\mathrm{per\,cent}$ of dark matter.
We aim to demonstrate the feasibility of using optical flux ratios of lensed quasars to study dark matter substructure. 
With ongoing and upcoming wide-field surveys, an increasing number of strongly lensed quasars are expected to be discovered. 
For instance, the ground-based Rubin Observatory Legacy Survey of Space and Time (LSST) is expected to detect approximately 3,500 strongly lensed quasars \citep{2019ApJ...873..111I,2025OJAp....8E...8A}. 
In addition to LSST, space-based surveys such as \textit{Euclid} \citep{2011arXiv1110.3193L} and \textit{CSST} \citep{2024MNRAS.533.1960C} are also projected to identify large samples of such systems.
This forthcoming wealth of optical flux ratio data will provide a valuable opportunity to probe the nature of dark matter using statistical analyses of flux ratio anomalies.
By simultaneously accounting for the effects of both dark matter substructure and stellar microlensing on optical flux ratios, we show that these measurements can be used to probe the properties of dark matter substructure and, consequently, place constraints on the fundamental nature of dark matter.
In this work, we focus on two types of dark matter substructures with distinct characteristics: (1) cold dark matter (CDM) subhalos and (2) fluctuations induced by fuzzy dark matter (FDM). In our study, we use the best fit results \citep{2020A&A...641A...6P} obtained by Planck in 2018 ($H_0=67.7\mathrm{\,km\,s^{-1}\,Mpc^{-1}}$, $\Omega_\mathrm{m}=0.31$).

The structure of this paper is as follows. 
Section \ref{2} describes the parameters of the strong lensing systems adopted in this work, as well as the method used to incorporate dark matter substructure. 
Section \ref{3} outlines the procedures for modeling and combining various perturbative effects, along with the statistical methods employed. 
In Section \ref{4}, we present the simulation results and provide a detailed discussion. 
Finally, Section \ref{5} summarizes our conclusions and offers perspectives for future work.

\begin{figure}
	\centering
	\begin{minipage}{\linewidth}
		\centering
		\includegraphics[width=\linewidth]{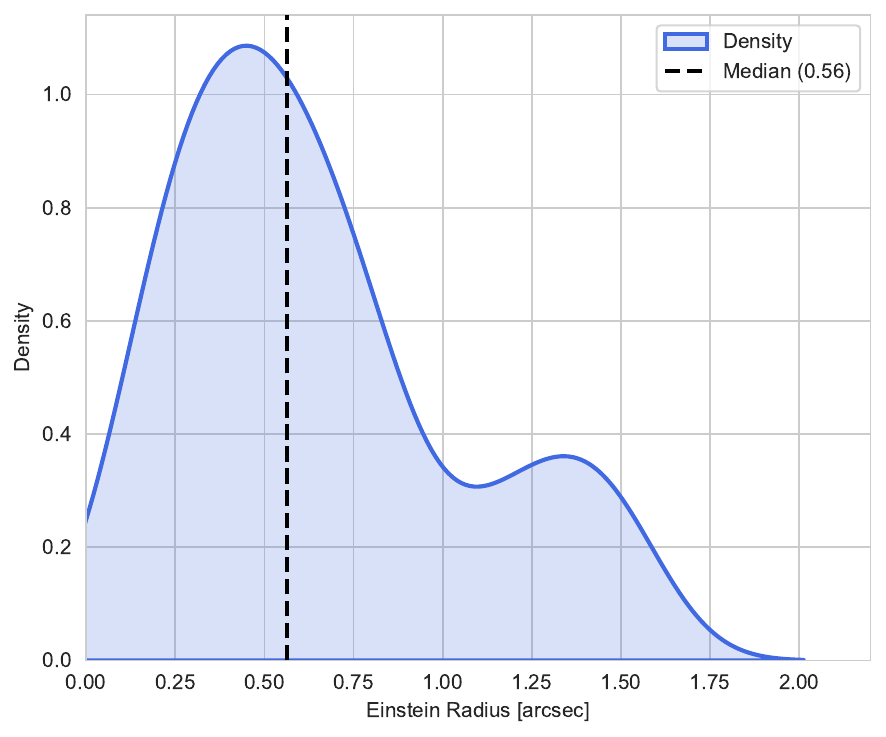}	
	\end{minipage}
    \caption{Probability density curve of the Einstein radii for all the selected strong lensing systems, obtained with a Gaussian kernel density estimator. The black dashed line indicates the median Einstein radius.}
    \label{ein_r.pdf}
\end{figure}

\begin{figure}
	\centering
	\begin{minipage}{\linewidth}
		\centering
		\includegraphics[width=\linewidth]{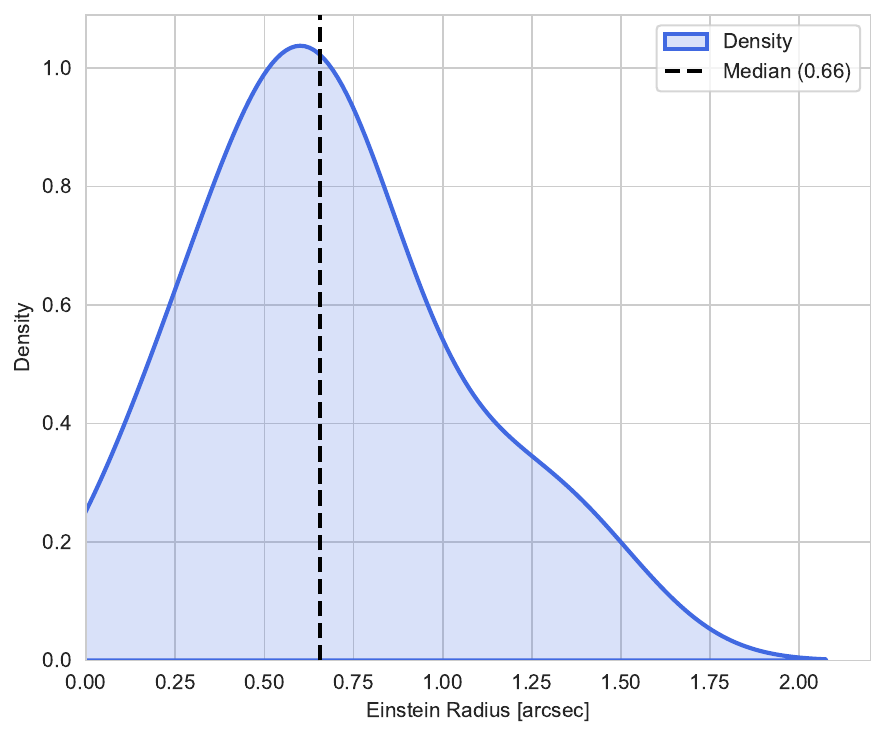}	
	\end{minipage}
    \caption{Probability density curve of the Einstein radii for the selected quads, obtained with a Gaussian kernel density estimator. The black dashed line indicates the median Einstein radius.}
    \label{ein_r_quad.pdf}
\end{figure}

\begin{figure}
	\centering
	\begin{minipage}{\linewidth}
		\centering
		\includegraphics[width=\linewidth]{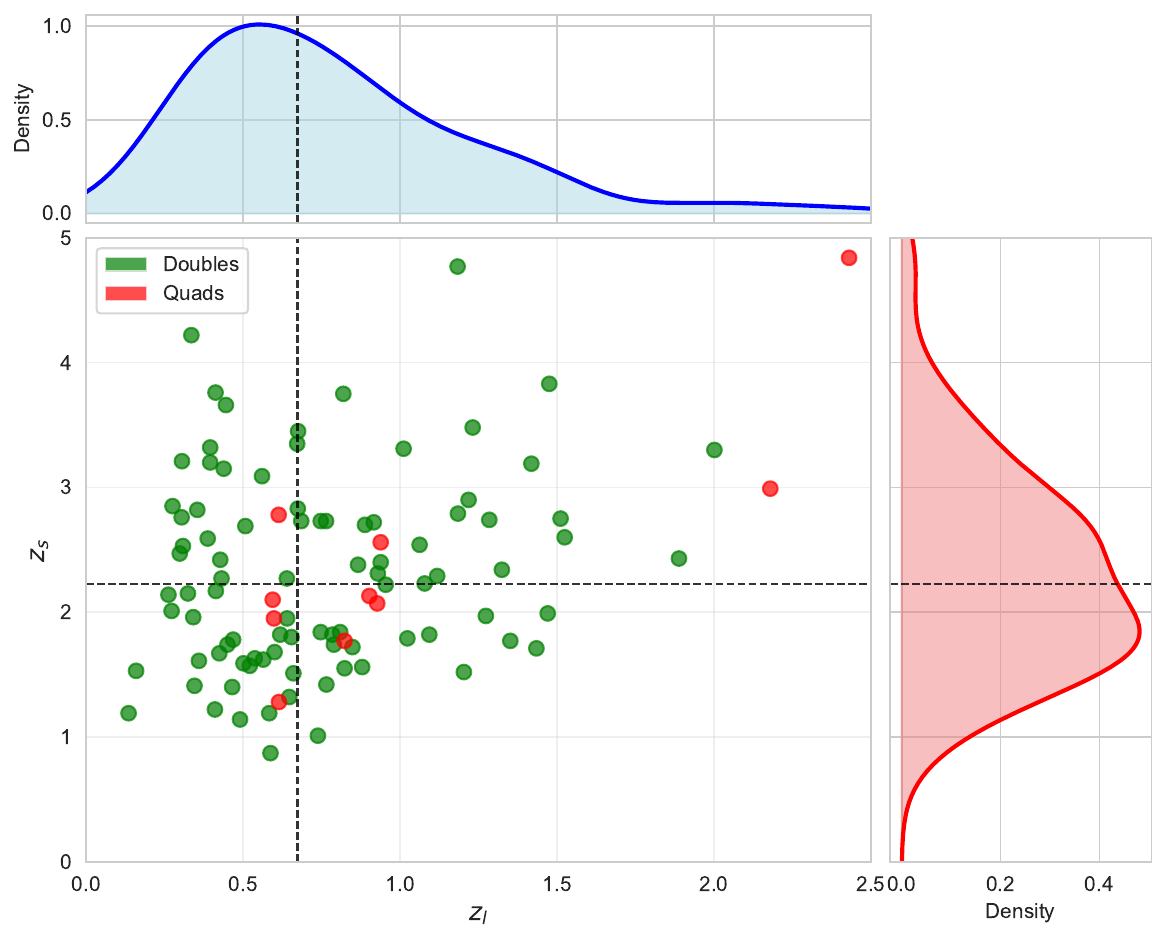}	
	\end{minipage}
    \caption{Redshift distribution of all strong lensing systems, with green and red dots representing doubles and quads, respectively. The black dashed lines mark the median lens and source redshifts. The upper and right panels show the probability density estimates of the lens and source redshift distributions, obtained with a Gaussian kernel density estimator.}
    \label{z_l_s.pdf}
\end{figure}

\section{Strong lensing systems}\label{2}
The strong lensing system we use consists of a quasar and a foreground galaxy. 
The lens mass model of the foreground galaxy includes a smooth macroscopic component as well as dark matter substructures whose properties are correlated with the parameters of the smooth lens model.
In Section \ref{2.1}, we introduce the parameters of the smooth lens model and the quasar. 
In Section \ref{2.2}, we describe the method used to incorporate CDM subhalos. 
In Section \ref{2.3}, we detail the procedure for adding FDM-induced fluctuations.

\subsection{Parameters of strong lensing systems}\label{2.1}
We select 100 strong lensing systems including 90 double-image systems (doubles) and 10 quadruple-image systems (quads) from the LSST mock catalog of gravitationally lensed quasars \footnote{\url{https://github.com/LSSTDESC/SL-Hammocks}} generated in \cite{2025OJAp....8E...8A} as the subjects of our study.
Fig. \ref{ein_r.pdf} and \ref{ein_r_quad.pdf} show the distributions of the Einstein radii $\theta_\mathrm{E}$ for all strong lensing systems and for the quads, respectively. Fig. \ref{z_l_s.pdf} shows the redshift distribution of all strong lensing systems.
The smooth lens model consists of an elliptical Navarro-Frenk-White (NFW) profile \citep{1996ApJ...462..563N,2021PASP..133g4504O} to describe the dark matter component, an elliptical Hernquist profile \citep{1990ApJ...356..359H,2021PASP..133g4504O} to represent the baryonic component (assuming the Salpeter initial mass function \citep{1955ApJ...121..161S}), and external shear. 
The position of the quasar in the source plane can be directly read from the mock catalog. 
However, since we later introduce perturbations from dark matter substructures, the image positions are not taken directly from the mock catalog but are instead calculated by solving the lens equation using \texttt{lenstronomy}
\footnote{\url{https://github.com/lenstronomy/lenstronomy}} \citep{2018PDU....22..189B,2021JOSS....6.3283B}, treating the quasar as a point source.

In the mock catalog, some strong lensing systems exhibit large Einstein radii. 
For such systems, computing the perturbations induced by substructures, especially those FDM-induced fluctuations, is computationally intensive and time-consuming.
However, since these systems constitute only a small fraction of the total sample, excluding them does not affect the overall conclusions of our study.
We also note that, within our selected samples, a small subset of systems with relatively small Einstein radii would likely be unobservable by LSST due to limitations imposed by its PSF. 
Nevertheless, these systems are still highly valuable for studying the nature of dark matter. 
For instance, low-mass dark matter halos are expected to produce significant FDM-induced fluctuations, which can place strong constraints on the FDM particle mass. 
Moreover, ongoing and upcoming wide-field surveys, such as those to be conducted by \textit{Euclid} and \textit{CSST}, are expected to discover a large number of strongly lensed quasars, and their higher angular resolution will allow them to detect these small Einstein radius systems. For this reason, we do not exclude them from our analysis.
Additionally, we adopt the LSST-based mock catalog because the underlying smooth lens mass models in this dataset explicitly separate the dark matter and baryonic components, making it convenient for incorporating stellar microlensing and FDM-induced fluctuations in our simulations.

\begin{figure}
	\centering
	\begin{minipage}{\linewidth}
		\centering
		\includegraphics[width=\linewidth]{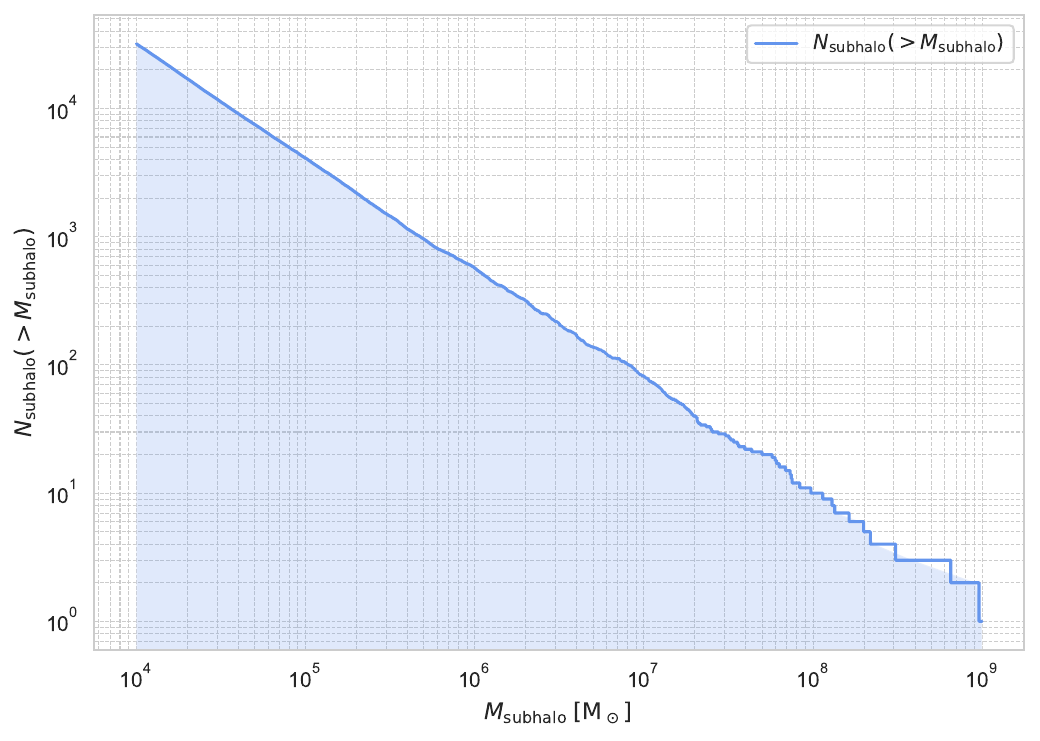}	
	\end{minipage}
    \caption{Cumulative number of CDM subhalos generated by \texttt{SASHIMI-C} for a strong lensing system with a host halo mass of $M_{200} = 10^{12.35}\,\mathrm{M}_\odot$, concentration parameter $c_{200} = 9.5$, lens redshift $z_l = 0.789$, and source redshift $z_s = 1.74$.}
    \label{num_subhalo_12.35.pdf}
\end{figure}

\begin{figure*}
	\centering
	\begin{minipage}{\linewidth}
		\centering
		\includegraphics[width=\linewidth]{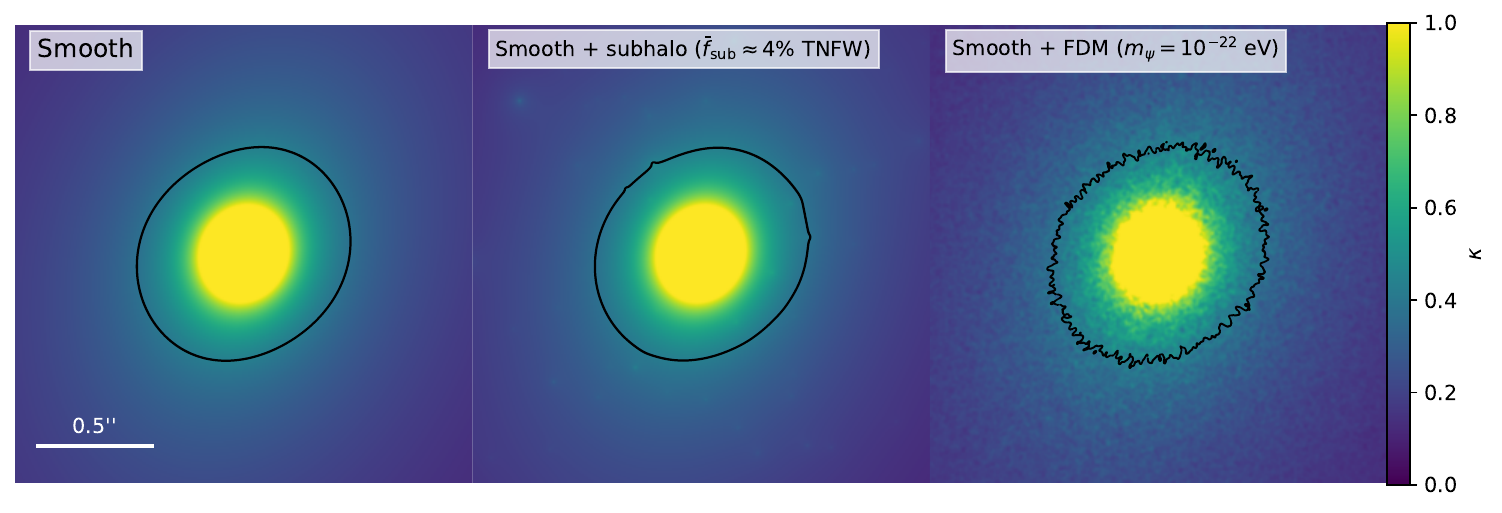}	
	\end{minipage}
    \caption{Surface mass density and critical curve maps for a strong lensing system with a host dark matter halo of $M_{200} = 10^{12.35}\,\mathrm{M}_\odot$, $c_{200} = 9.5$, lens redshift $z_l = 0.789$, and source redshift $z_s = 1.74$. The left panel shows the smooth mass model without any substructure. The middle panel includes CDM subhalos with a mean substructure mass fraction of $\bar{f}_\mathrm{sub} \approx 0.04$, modeled with a TNFW profile. The right panel includes FDM-induced fluctuations corresponding to an ultra-light boson mass of $m_{\psi} = 10^{-22}\,{\mathrm{eV}}$.}
    \label{kappa_and_mag_1e710_14.pdf}
\end{figure*}

\subsection{CDM subhalos}\label{2.2}
CDM subhalos within dark matter halos exhibit different properties depending on their mass and redshift. 
To account for this diversity, we generate CDM subhalos using the Semi-Analytical SubHalo Inference ModelIng for CDM (\texttt{SASHIMI-C}) \footnote{\url{https://github.com/shinichiroando/sashimi-c}} from \cite{2018PhRvD..97l3002H, 2020JPhCS1468a2050H}. 
\texttt{SASHIMI-C} can be used to efficiently compute the subhalo mass function and provides a complete catalog of CDM subhalos within a host halo of arbitrary mass and redshift.
To ensure consistency with \cite{2025OJAp....8E...8A}, we adopt the mass-concentration relation presented by \cite{2019ApJ...871..168D} and use the \texttt{COLOSSUS} package \citep{2018ApJS..239...35D}.  
We model the subhalos as tidally truncated Navarro-Frenk-White (TNFW) profile \citep{2009JCAP...01..015B}

\begin{equation}
\rho(r, r_s, r_t) = \frac{\rho_s}{(r/r_s)(1 + r/r_s)^2} \cdot \frac{r_t^2}{r^2 + r_t^2},
\label{tnfw}
\end{equation}
where $r_t$ is a truncation radius and $r_s$ is the NFW profile scale radius.

Although we adopt a different truncation method than that used in \texttt{SASHIMI-C}, the central density dominates the flux ratio signal. 
As a result, different choices for the truncation have little impact on our conclusions. 
Unless otherwise stated, we consider subhalos with masses within the range $10^7 \le M_\mathrm{subhalo}/\mathrm{M}_\odot \le 10^{10}$. 
Subhalos more massive than $ 10^{10}\,\mathrm{M}_\odot$ would likely contain a luminous galaxy. 
Since our goal is to propose a new method for detecting dark matter substructure, we prioritize computational efficiency without compromising the validity of our conclusions.
Therefore, we neglect CDM subhalos with masses below $10^{7}\,\mathrm{M}_\odot$, as their impact on flux ratios is relatively small. 
As demonstrated in Section \ref{4}, excluding low-mass subhalos does not alter our main conclusions.
This simplification is further supported by the fact that the number of subhalos increases rapidly with decreasing $ M_\mathrm{subhalo}$ (see Fig. \ref{num_subhalo_12.35.pdf}).
Once the CDM subhalos are generated, they are uniformly and randomly distributed within $3\theta_\mathrm{E}$ \citep{2015MNRAS.447.3189X}.
To conserve the total mass within $\theta_\mathrm{E}$, we additionally include a negative mass sheet.

To explore the feasibility of studying dark matter substructures through optical flux ratios, in addition to the CDM subhalo generation method described above, we modify the CDM subhalo model in two ways to account for discrepancies between theory and observations. 
First, considering the uncertainty in the internal density profiles of the subhalo, we treat subhalos as point masses. 
Second, given that the CDM subhalo mass fraction $\bar{f}_\mathrm{sub} \approx 0.04$ within $3\theta_\mathrm{E}$, we also test a scenario where this fraction is doubled to 
$\bar{f}_\mathrm{sub} \approx 0.08$, in order to account for the uncertainty in substructure abundance. 

\subsection{FDM-induced fluctuations}\label{2.3}
Fuzzy dark matter (FDM) is a dark matter candidate composed of ultra-light bosons with $m_{\psi} \sim 10^{-22}\,{\mathrm{eV}}$. 
Due to their extremely large de Broglie wavelengths, which are much greater than the average inter-particle spacing, FDM can be well described as a classical wave phenomenon \citep{2021ARA&A..59..247H}. 
The density of FDM halos fluctuates on the scale of the de Broglie wavelength $\lambda_\mathrm{dB}$, with values ranging from zero to twice the local mean density, corresponding to destructive and constructive interference, respectively. 
These surface density fluctuations can be modeled as a Gaussian random field (GRF), with the fluctuation profile well approximated by a two-dimensional Gaussian function. 
The full width at half maximum (FWHM) of this Gaussian is equal to the de Broglie wavelength $\lambda_\mathrm{dB}$ \citep{2020PhRvL.125k1102C,2023NatAs...7..736A,2024A&A...690A.359D}.
For  a given halo mass $M_h$, $\lambda _{\mathrm{dB}}$ is set by the boson mass, $m_{\psi }$ according to the relationship \citep{2014PhRvL.113z1302S,2016ApJ...818...89S}:

\begin{equation}
\label{equation:lambda}
    \lambda _{\mathrm{dB}}=150\left (\frac{10^{-22}\,{\mathrm{eV}}}{m_{\psi }}\right )\left ( \frac{M_{h}}{10^{12}\,\mathrm{M}_\odot} \right )^{-1/3}{\mathrm{pc}}.
\end{equation}
The variance of the surface mass density at each point on the dark matter halo can be expressed as \citep{2020PhRvL.125k1102C,2023NatAs...7..736A}:

\begin{equation}
\label{equation:sigma}
\sigma_\mathrm{\Sigma}^2(\boldsymbol{\xi})=\lambda_{\mathrm{dB}}\sqrt{\pi}\int_{-\infty}^\infty\rho_\mathrm{DM}^2(z,\boldsymbol{\xi})dz,
\end{equation}
where $z$ is the coordinate of the $z$-axis along the line-of-sight direction, $\boldsymbol{\xi}$ is the projection radius vector from the center of the halo and $\rho_\mathrm{DM}$ is the smooth density profile of the dark matter component.

To generate a dark matter halo composed of FDM, we first create a white noise field modulated by the variance given in Equation (\ref{equation:sigma}). 
We then convolve this field with a two-dimensional Gaussian kernel with full width at half maximum (FWHM) equal to $\lambda _{\mathrm{dB}}$, in order to introduce correlations \citep{2023MNRAS.524L..84P}. 
We set $m_{\psi} = 10^{-22}\,{\mathrm{eV}}$. 
To explore the feasibility of studying dark matter substructures through optical flux ratios, we also consider the case with $m_{\psi} = 5 \times 10^{-22}\,{\mathrm{eV}}$, which corresponds to a smaller $\lambda _{\mathrm{dB}}$ and hence results in weaker perturbations.
Fig. \ref{kappa_and_mag_1e710_14.pdf} shows the surface mass density and critical curves for the smooth mass model, as well as after the addition of perturbations.

\begin{figure*}
	\centering
	\begin{minipage}{\linewidth}
		\centering
		\includegraphics[width=\linewidth]{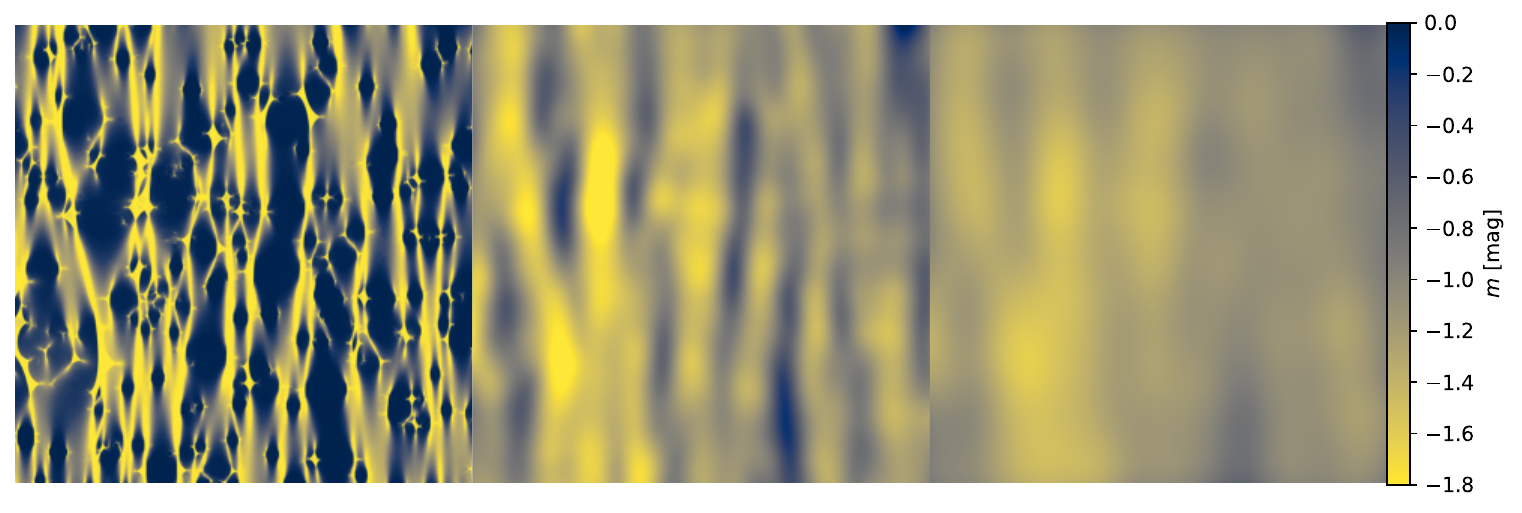}	
	\end{minipage}
    \caption{Microlensing simulation for an image with total convergence $\kappa = 0.511$, stellar convergence $\kappa_* = 0.112$, and shear $\gamma = 0.748$, with lens and source redshifts $z_l = 0.789$ and $z_s = 1.74$, respectively. The left panel shows the magnification map. The middle and right panels show the convolved maps after applying a Gaussian profile corresponding to an accretion disk with a half-light radius of 3 light-days and 6 light-days, respectively.}
    \label{ml_plot.pdf}
\end{figure*}

\begin{figure}
	\centering
	\begin{minipage}{\linewidth}
		\centering
		\includegraphics[width=\linewidth]{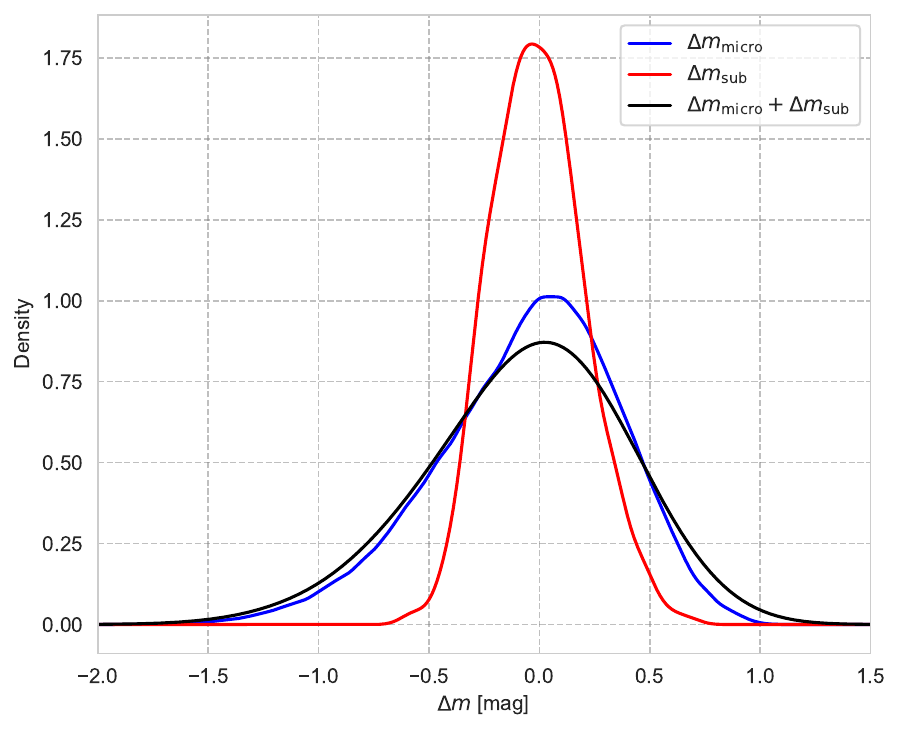}	
	\end{minipage}
    \caption{The probability density distributions of $\Delta m_\mathrm{micro}$, obtained using an accretion disk with a half-light radius of 3 light-days, and $\Delta m_\mathrm{sub}$, corresponding to FDM-induced fluctuations with an ultra-light boson mass of $m_{\psi} = 10^{-22}\,{\mathrm{eV}}$, are shown for a strong lensing system with a dark matter halo of $M_{200} = 10^{12.35}\,\mathrm{M}_\odot$, concentration parameter $c_{200} = 9.5$, lens redshift $z_l = 0.789$, and source redshift $z_s = 1.74$. Also shown is the resulting probability density distribution of $\Delta m_\mathrm{micro} + \Delta m_\mathrm{sub}$, obtained by convolving the two individual distributions. The blue, red, and black curves correspond to $\Delta m_\mathrm{micro}$, $\Delta m_\mathrm{sub}$, and their convolution, respectively.}
    \label{con_noerror.pdf}
\end{figure}

\section{Methods}\label{3}
To ensure the statistical independence of the flux ratios, we select the flux ratio between the minimum and saddle point images, i.e., $\mu_\mathrm{minimum}/\mu_\mathrm{saddle}$, as the subject of our study. 
This selection results in one independent flux ratio for each double and two for each quad.
Flux ratio perturbations in the optical band, i.e. flux ratio anomalies, can generally be categorized into three components: perturbations induced by dark matter substructures, perturbations due to microlensing, and errors from observation and modeling.

In Section \ref{3.1}, we introduce substructure-induced perturbations. In Section \ref{3.2}, we describe microlensing-induced perturbations. Section \ref{3.3} discusses the total flux ratio perturbations. In Section \ref{3.4} , we present our statistical methodology.

\subsection{Substructure-induced perturbations}\label{3.1}
To quantify the impact of dark matter substructure on flux ratios, we fix the source position and the parameters of the smooth lens model for each system. 
For each dark matter substructure scenario, we generate 1000 realizations to obtain the distribution of flux ratio perturbations induced by dark matter substructure.
During this process, extra images may occasionally occur; however, such cases lie beyond the scope of this work and are therefore neglected.

In principle, under the FDM scenario, one should consider both FDM-induced fluctuations and the contribution from FDM subhalos.
Additionally, line-of-sight structures can also contribute to flux ratio anomalies, and in realistic analyses they must be taken into account \citep{2012MNRAS.421.2553X}.
Due to the suppression of small-scale structure formation by quantum pressure in FDM, the abundance of low-mass halos is significantly reduced and their concentration parameters differ from the CDM case \citep{2014NatPh..10..496S,2016ApJ...818...89S,2017MNRAS.465..941D,2022MNRAS.510.1425K}.
However, since the aim of this work is to propose and demonstrate a methodological framework for utilizing optical flux ratios, we intentionally simplify our model by neglecting line-of-sight structures in both CDM and FDM scenarios, and additionally neglecting FDM subhalos in the FDM scenario. 
This allows us to focus on the interplay between the two primary types of substructure considered in this study (CDM subhalos for the CDM scenario and FDM-induced fluctuations for the FDM scenario) and stellar microlensing.
This choice allows for a clearer assessment of the fundamental feasibility of our method without introducing additional computational complexities and degeneracies. We acknowledge that the inclusion of line-of-sight structures would enhance the total perturbation signal in both scenarios, while the inclusion of FDM subhalos would further enhance the signal in the FDM scenario, potentially reducing the number of flux ratio measurements required for detecting dark matter substructure.
A detailed, combined analysis incorporating line-of-sight structures and FDM subhalos (where applicable) is an important next step, but it lies beyond the scope of this foundational work and is deferred to a future study.
Meanwhile, the central region of a dark matter halo composed of FDM is expected to host a soliton core, whose characteristic scale is comparable to the de Broglie wavelength \citep{2014NatPh..10..496S}. 
Since the flux ratio anomalies are primarily induced by density fluctuations near the critical curve, we neglect the central soliton and instead adopt the NFW profile directly.

\subsection{Microlensing-induced perturbations}\label{3.2}
When considering the effects of microlensing on flux ratios, we use the smooth mass model and neglect the influence of substructures (a justification for this simplification is provided later).
We use \texttt{MULES} \footnote{\url{https://github.com/gdobler/mules}} to simulate the magnification map. 
Based on the smooth mass model and the quasar position derived earlier, we solve the lens equation to obtain the total convergence $\kappa$, stellar convergence $\kappa_*$, and total shear $\gamma$ at each image position.

The stellar mass is assumed to follow the Salpeter initial mass function \citep{1955ApJ...121..161S}, with a mean stellar mass of $\langle M_* \rangle = 0.2\,\mathrm{M}_\odot$ and an upper-to-lower mass ratio of $M_\mathrm{upper}/M_\mathrm{lower} = 6.46\,\mathrm{M}_\odot / 0.06\,\mathrm{M}_\odot = 100$ \citep{2021A&A...647A.115C}.
The size of the magnification map is scaled to the Einstein radius of $\left \langle M_* \right \rangle$, $R_E$, on the source plane, described as:

\begin{equation}
\label{equation:re}
R_E = D_{\text{S}} \times \sqrt{\frac{4G\langle M_* \rangle}{c^2} \frac{D_{\text{LS}}}{D_{\text{L}} D_{\text{S}}}},
\end{equation}
where $D_\mathrm{L}$, $D_\mathrm{S}$, and $D_\mathrm{LS}$ are angular diameter distances to the lensing galaxy (deflector) located at redshift $z_l$, to the source located at redshift $z_s$ and between them, respectively.

The flux is obtained by convolving the accretion disk's light profile with the magnification map. 
We model the light profile of the accretion disk as a Gaussian distribution and perform the convolution over a region that contains 99 $\mathrm{per\,cent}$ of the total flux. 
However, the specific form of the light profile has only a minor impact on microlensing, whereas the size of the accretion disk, i.e., its half-light radius $R_{1/2}$, plays a dominant role \citep{2005ApJ...628..594M}.

In a standard, non-relativistic, thin-disk model that emits as a blackbody, the characteristic radius of the disk at rest-frame wavelength $\lambda_\mathrm{rest}$ is given by \citep{1973A&A....24..337S}:

\begin{equation}
\label{equation:rs}
R_s = 9.7 \times 10^{15} \ \text{cm} \left( \frac{\lambda_{\text{rest}}}{\mu\text{m}} \right)^{4/3} \left( \frac{M_{\text{BH}}}{10^9\,\mathrm{M}_\odot} \right)^{2/3} \left( \frac{L}{\eta L_E} \right)^{1/3},
\end{equation}
where $M_\mathrm{BH}$ is the central black hole mass, $L/L_E$ is luminosity in units of the Eddington luminosity, $\eta$ is the accretion efficiency and $R_{1/2} = 2.44R_s$.
We set $\bar{M}_\mathrm{BH} = 5.5 \times 10^8\,\mathrm{M}_\odot$ \citep{2018ApJ...862..123M}. Given that our lens samples are drawn from a mock catalog in the i-band and $L/L_E \sim 1/3$, $\eta \sim 0.1$ \citep{2006ApJ...648..128K,2009ApJ...698.1550H,2010A&A...516A..87S}. 
Based on the thin-disk model, we fix the half-light radius to 3 light-days.
We also note that accretion disk sizes measured by methods such as continuum  reverberation mapping are larger than those predicted by the thin-disk model \citep{2010ApJ...712.1129M,2018ApJ...862..123M,2020ApJ...895..125C,2023A&A...673A..88A}. 
To reflect for this discrepancy, we also adopt a half-light radius that is twice the thin-disk prediction, corresponding to 6 light-days.  
Although we fix the size of the accretion disk in our analysis, the effect of varying the disk size on microlensing, through changes in the ratio $R_{1/2}/R_E$, is degenerate with the effect of changing the redshift of the strong lensing system. In other words, varying the redshift is equivalent to varying $R_{1/2}$ in terms of its impact on microlensing \citep{2019MNRAS.483.5583V}.

After accounting for the reduction in magnification map size due to convolution with the source light profile, we generate final convolved maps with the size of $40 \times R_E$ in 3360 pixels for each of the two half-light radii considered (see Fig. \ref{ml_plot.pdf}). 
For each map, we randomly sample 1000 points to obtain fluxes. 
Consequently, for each flux ratio, we obtain a distribution of microlensing-induced perturbations with $10^6$ samples.

\subsection{Total flux ratios}\label{3.3}
For a strong lensing system, the observed flux ratio in the optical band can be expressed in magnitudes as:

\begin{equation}
m = m_\mathrm{macro} + \Delta m_\mathrm{micro} + \Delta m_\mathrm{sub} + \Delta m_\mathrm{SL} + \Delta m_\mathrm{obs},
\label{mag1}
\end{equation}

\begin{equation}
m = -2.5 \log_{10} \left( \frac{\mu_\mathrm{minimum}}{\mu_\mathrm{saddle}} \right),
\label{mag0}
\end{equation}
here, $m$ denotes the observed flux ratio, $m_\mathrm{macro}$ is the flux ratio predicted by the smooth macroscopic lens model, $\Delta m_\mathrm{micro}$ represents the microlensing-induced perturbation, $\Delta m_\mathrm{sub}$ corresponds to the substructure-induced perturbation, $ \Delta m_\mathrm{SL}$ accounts for errors from the smooth lens model fitting, and $\Delta m_\mathrm{obs}$ is the observational error.

In principle, $\Delta m_\mathrm{micro}$ and $\Delta m_\mathrm{sub}$ are correlated, since dark matter substructures perturb $\kappa$ and $\gamma$, which can affect the microlensing. 
However, all of the above errors and perturbations can be regarded as first-order deviations from the flux ratio predicted by the smooth model. 
The influence of substructures on microlensing is a higher-order effect and thus is neglected. 
We therefore assume these components contribute independently to the total flux ratio perturbation.

For each individual flux, we assume $\sigma_\mathrm{SL} \approx 0.1\, \mathrm{mag}$ \citep{2023MNRAS.518.1260S} and $\sigma_\mathrm{obs} \approx 0.1\, \mathrm{mag}$ (corresponding to a 10 $\mathrm{per\,cent}$ flux measurement uncertainty). 
We conservatively estimate $\sigma_\mathrm{SL}$ based on the modeling results for quads presented in \cite{2023MNRAS.518.1260S}. 
However, the actual value of $\sigma_\mathrm{SL}$ for doubles may be larger. 
As we will demonstrate later, even if we have underestimated $\sigma_\mathrm{SL}$ for doubles, the constraints obtained using only quads still provide stronger constraints than those from all strong lensing systems or those from doubles alone. 
If doubles indeed have a larger $\sigma_\mathrm{SL}$, this would only strengthen our conclusions rather than alter them.
Using the error propagation formula (assuming normally distributed errors), we obtain

\begin{equation}
\Delta m_\mathrm{error} = \Delta m_\mathrm{SL} +\Delta m_\mathrm{obs},
\label{mag2}
\end{equation}

\begin{equation}
\sigma_\mathrm{error} = \sqrt{2(\sigma_\mathrm{SL}^2 + \sigma_\mathrm{obs}^2)} \approx 0.2\,\mathrm{mag},
\label{mag3}
\end{equation}

\begin{equation}
\Delta m = m - m_\mathrm{macro} = \Delta m_\mathrm{micro} + \Delta m_\mathrm{sub} + \Delta m_\mathrm{error},
\label{mag4}
\end{equation}
here, $\Delta m_\mathrm{error}$ represents the observational uncertainties, which combine the errors from the smooth lens model fitting and the observational measurement noise. 
The quantity $\sigma_\mathrm{error}$ denotes the standard deviation of $\Delta m_\mathrm{error}$. It should be noted that $\sigma_\mathrm{error}$ refers to the standard deviation of the flux ratio, rather than the standard deviations of the individual flux ($\sigma_\mathrm{SL}$ and $\sigma_\mathrm{obs}$) introduced earlier.
Since the probability density functions (PDFs) of $\Delta m_\mathrm{micro}$ and $\Delta m_\mathrm{sub}$ do not have analytical forms, combining these independent perturbations requires convolving their respective PDFs. 
In Fig. \ref{con_noerror.pdf}, we present the resulting distribution of $\Delta m_\mathrm{micro} + \Delta m_\mathrm{sub}$ obtained by convolving the PDFs of $\Delta m_\mathrm{micro}$ and $\Delta m_\mathrm{sub}$. 
This approach enables us to account for multiple types of perturbations simultaneously and derive the combined probability density function that incorporates their joint effects.

It is worth noting that the intrinsic time variability of quasars can also have a significant impact on flux ratios. 
The measurement uncertainty of time delays by LSST is less than 3 $\mathrm{per\,cent}$ \citep{2015ApJ...800...11L}. 
Following the methodology of \cite{2008A&A...478...95Y}, and considering a time-delay measurement error of $\sim 1\, \mathrm{day}$ , observations in the i-band, and quasar magnitudes ranging from $-21$ to $-30$, the expected flux ratio variation due to intrinsic time variability is $< 0.05 \, \mathrm{mag}$ in the worst-case scenario. 
Furthermore, to further reduce the influence of intrinsic time variability, one can perform multiple flux measurements within a timescale of $\sim 1\, \mathrm{month}$ (during which microlensing effects remain nearly unchanged). 
Based on the measured time delays, observe the corresponding delayed fluxes within the same timescale and then take the average. 
However, the exact magnitude of the impact from intrinsic time variability depends on the specific observational strategy and algorithms adopted. 
Therefore, we neglect the effect of intrinsic time variability in this work, as doing so does not affect our conclusions.

\subsection{Statistical methodology}\label{3.4}
To demonstrate that flux ratios in the optical band can be used to probe dark matter substructure, we need to show that, given a sufficient number of flux ratio observations, the observable quantity $\Delta m$ and the computed quantity 
$\Delta m_\mathrm{micro} + \Delta m_\mathrm{error}$ do not originate from the same underlying model. 
Here, $\Delta m_\mathrm{micro} + \Delta m_\mathrm{error}$ represents flux ratio perturbations arising from microlensing and observational uncertainties only, while $\Delta m$ includes contributions from microlensing, dark matter substructure, and observational uncertainties.
Furthermore, to show that flux ratios in the optical band are sensitive to different dark matter substructure parameters, we need to demonstrate that, given a sufficient number of  flux ratios observations, the flux ratio perturbations $\Delta m$ corresponding to different dark matter substructure parameters do not arise from the same underlying model.

To address these requirements, we employ the Kolmogorov–Smirnov (KS) test. 
When the null hypothesis is true, the corresponding p-values are expected to be uniformly distributed between 0 and 1 \footnote{\url{https://statproofbook.github.io/P/pval-h0}}. 
For a two-tailed hypothesis, the p-value is defined as:

\begin{equation}
p = 2 \cdot \min \left( \left[ F_T(t_{\text{obs}}),\ 1 - F_T(t_{\text{obs}}) \right] \right),
\label{ks}
\end{equation}
where $F_T(t)$ is the cumulative distribution function (CDF) under the null hypothesis $H_0$ and $t_{\text{obs}}$ is the observed test statistic. 
In this work, the observed test statistic refers to the measured flux ratio perturbation, denoted as $\Delta m$. 
Specifically, the KS test quantifies the distance between the cumulative distribution function (CDF) of the measured p-values and the uniform distribution. 
This distance can be translated into a single p-value, $p_\mathrm{ks}$ (ranging from 0 to 1), which indicates how closely the two CDFs match.
If $p_\mathrm{ks} < 0.05$, we can conclude that the null hypothesis $H_0$ is rejected at the 95 $\mathrm{per\,cent}$ confidence level.

Therefore, by setting the null hypothesis $H_0$ such that the flux ratio perturbations arise solely from microlensing and observational uncertainties, and assuming that the observed statistic $\Delta m$ is drawn from a distribution that includes the effects of microlensing, dark matter substructure, and observational uncertainties, we can apply the Kolmogorov–Smirnov (KS) test to determine the number of observations required to reject the null hypothesis at a significance level of 95 $\mathrm{per\,cent}$. 
A rejection at this level would indicate that microlensing and observational uncertainties alone are insufficient to explain the observed optical flux ratio anomalies, thus demonstrating the detectability of dark matter substructure via optical flux ratios.
Specifically, using the 100 strong lensing systems we selected earlier, corresponding to 110 independent flux ratios, we treat the set of 110 flux ratios as a representative sample of the underlying flux ratio distribution.
Each 'observation' corresponds to first randomly selecting a flux ratio distribution, and then drawing a flux ratio value $m$ from the corresponding probability density function.
Then we compute the corresponding p-value from the null distribution $\Delta m_\mathrm{micro} + \Delta m_\mathrm{error}$. 
For a fixed number of such observations, we compute the KS test p-value, $p_\mathrm{ks}$, (note that this is distinct from the individual p-value) and repeat the procedure 100 times to obtain an average $p_\mathrm{ks}$.

Similarly, by setting the null hypothesis $H_0$ such that the flux ratio perturbations arise from microlensing, observational uncertainties, and a fiducial dark matter substructure model (e.g., CDM subhalos with a mean substructure mass fraction $\bar{f}_\mathrm{sub} \approx 0.04$ modeled with a TNFW profile, or FDM with $m_{\psi} = 10^{-22}\,{\mathrm{eV}}$), and assuming that the observed $\Delta m$ values are drawn from a model with different substructure parameters, we can determine the number of observations required to statistically distinguish between two dark matter scenarios. 
This further demonstrates that optical flux ratios can be used to probe the properties of dark matter substructure and constrain the underlying dark matter model parameters.

\begin{figure*}
	\centering
	\begin{minipage}{\linewidth}
		\centering
		\includegraphics[width=\linewidth]{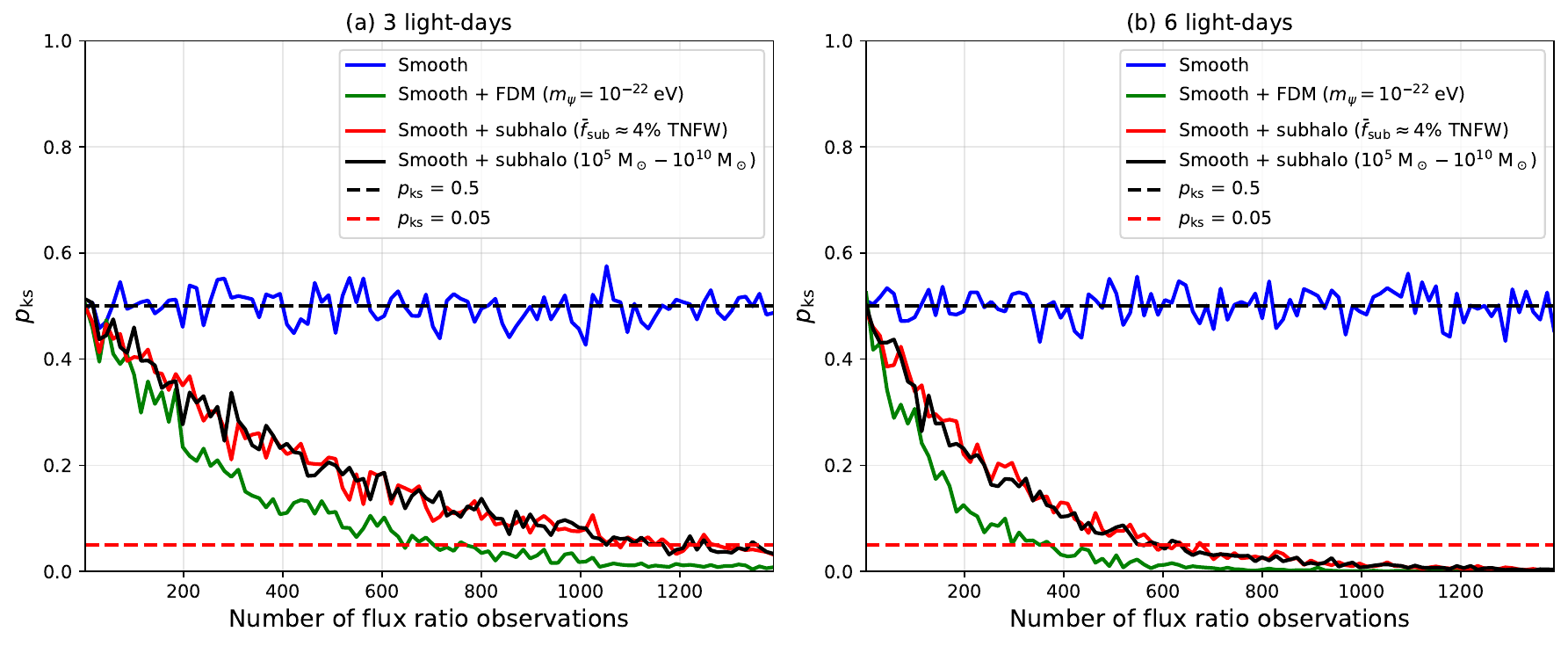}	
	\end{minipage}
    \caption{The horizontal axis in the plot represents the number of optical flux ratio observations, while the vertical axis corresponds to the associated KS test p-value ($p_\mathrm{ks}$). In this case, we  use flux ratios from all strong lensing systems. The null hypothesis ($H_0$, blue line) assumes perturbations arise solely from microlensing and observational uncertainties. Other lines show results when the observed data are drawn from models including CDM subhalos (red and black) or FDM-induced fluctuations (green). The dashed lines indicate $p_\mathrm{ks}=0.5$ and the 95 $\mathrm{per\,cent}$ rejection threshold of $p_\mathrm{ks}=0.05$. Left and right panels correspond to accretion disk half-light radii of 3 and 6 light-days, respectively.}
    \label{ks_test_base_all.pdf}
\end{figure*}

\begin{figure*}
	\centering
	\begin{minipage}{\linewidth}
		\centering
		\includegraphics[width=\linewidth]{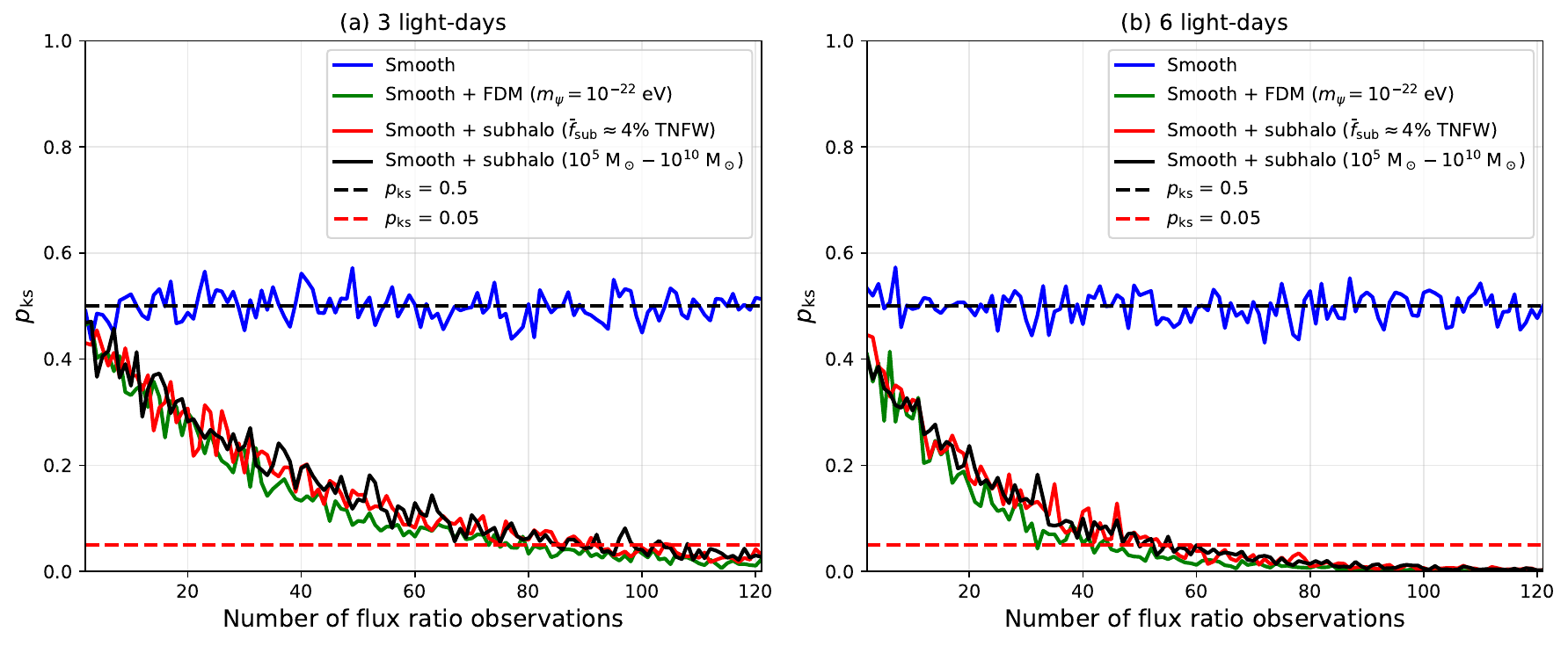}	
	\end{minipage}
    \caption{The horizontal axis in the plot represents the number of optical flux ratio observations, while the vertical axis corresponds to the associated KS test p-value ($p_\mathrm{ks}$). In this case, we only use flux ratios from quads. The null hypothesis ($H_0$, blue line) assumes perturbations arise solely from microlensing and observational uncertainties. Other lines show results when the observed data are drawn from models including CDM subhalos (red and black) or FDM-induced fluctuations (green). The dashed lines indicate $p_\mathrm{ks}=0.5$ and the 95 $\mathrm{per\,cent}$ rejection threshold of $p_\mathrm{ks}=0.05$. Left and right panels correspond to accretion disk half-light radii of 3 and 6 light-days, respectively.}
    \label{ks_test_base_quad.pdf}
\end{figure*}

\begin{figure*}
	\centering
	\begin{minipage}{\linewidth}
		\centering
		\includegraphics[width=\linewidth]{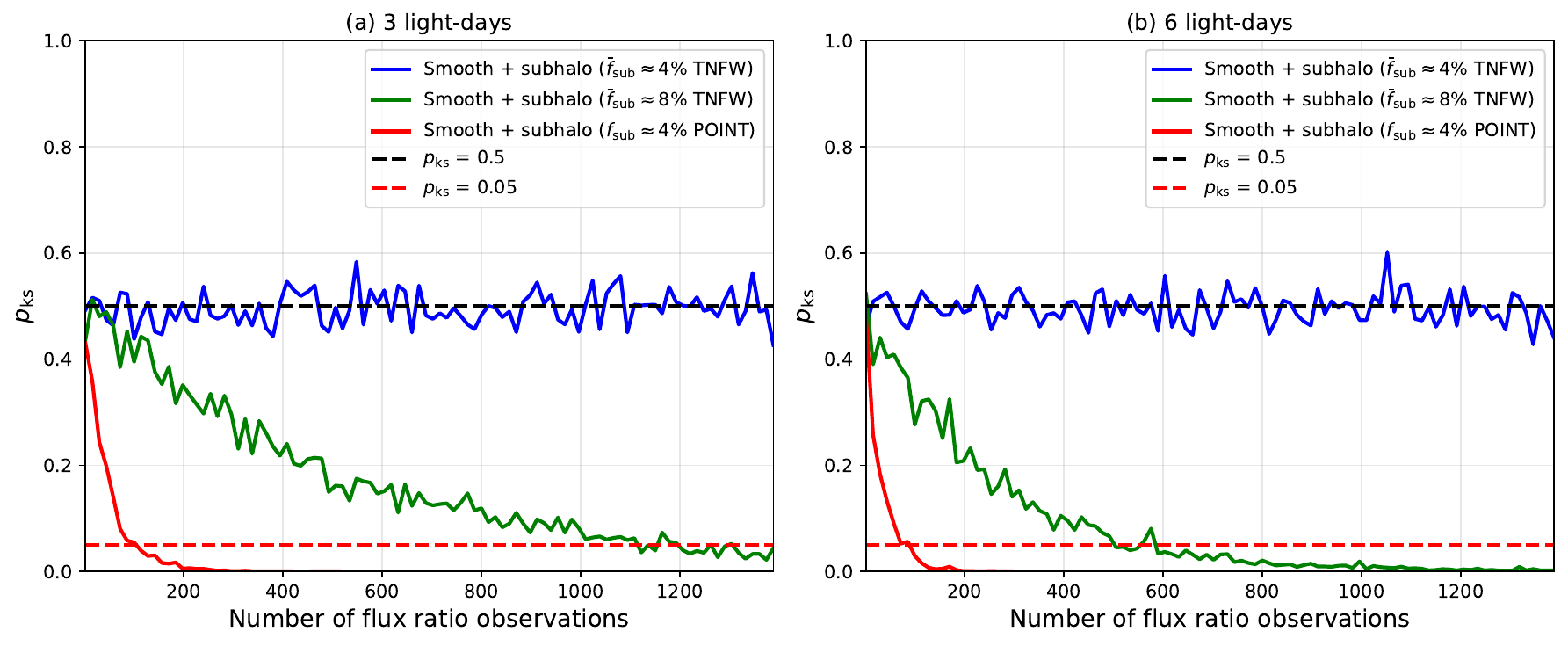}	
	\end{minipage}
    \caption{The horizontal axis in the plot represents the number of optical flux ratio observations, while the vertical axis corresponds to the associated KS test p-value ($p_\mathrm{ks}$). Here, we only use flux ratios from quads. The null hypothesis ($H_0$, blue line) is a fiducial CDM subhalo model ($\bar{f}_{\rm sub}\approx4\%$ TNFW). The alternative models are a CDM model with doubled substructure mass fraction (green) and a model where subhalos are point masses (red). The dashed lines indicate $p_\mathrm{ks}=0.5$ and the 95 $\mathrm{per\,cent}$ rejection threshold of $p_\mathrm{ks}=0.05$. Left and right panels correspond to accretion disk half-light radii of 3 and 6 light-days, respectively.}
    \label{ks_test_subhalo_quad.pdf}
\end{figure*}

\begin{figure*}
	\centering
	\begin{minipage}{\linewidth}
		\centering
		\includegraphics[width=\linewidth]{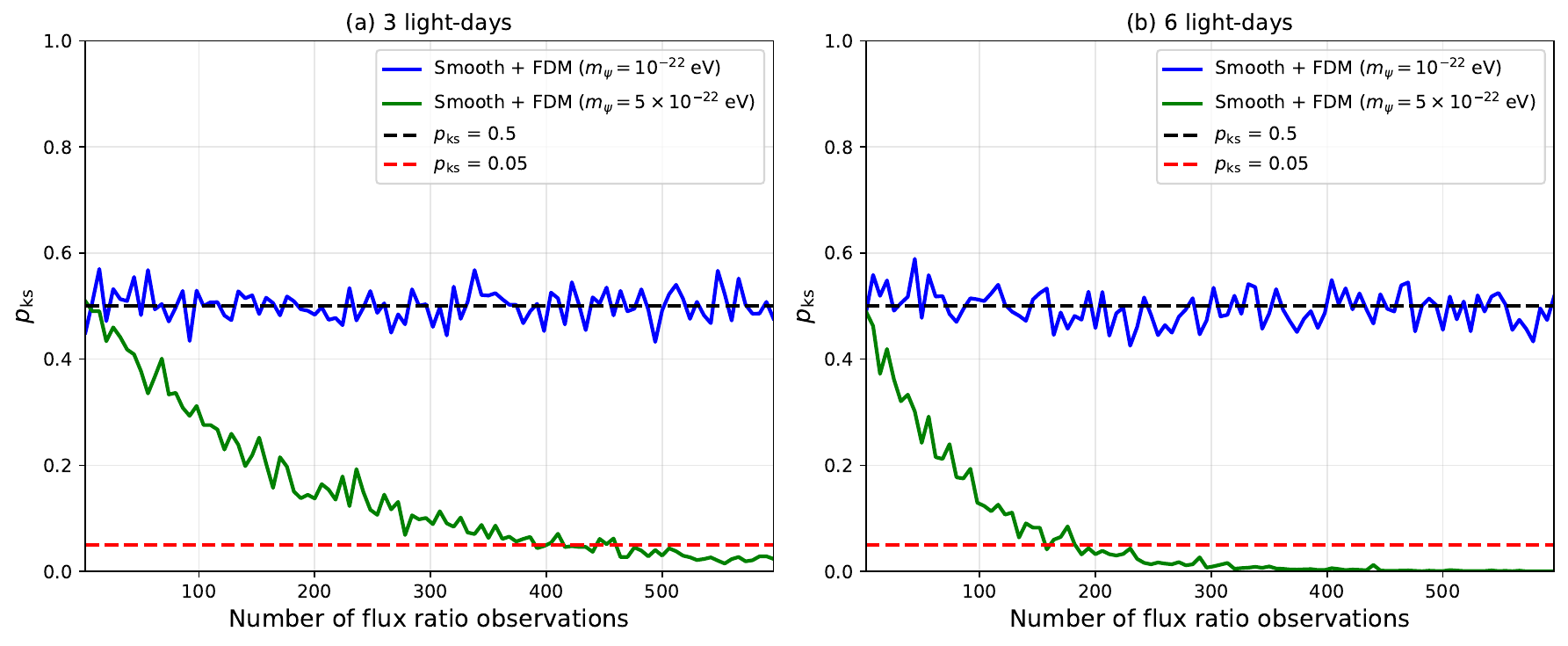}	
	\end{minipage}
    \caption{The horizontal axis in the plot represents the number of optical flux ratio observations, while the vertical axis corresponds to the associated KS test p-value ($p_\mathrm{ks}$). In this plot, we only use flux ratios from quads. The null hypothesis ($H_0$, blue line) corresponds to an FDM model with $m_{\psi}=10^{-22}\,\mathrm{eV}$. The alternative model (green) assumes a higher ultra-light boson mass of $m_{\psi}=5\times10^{-22}\,\mathrm{eV}$. The dashed lines indicate $p_\mathrm{ks}=0.5$ and the 95 $\mathrm{per\,cent}$ rejection threshold of $p_\mathrm{ks}=0.05$. Left and right panels correspond to accretion disk half-light radii of 3 and 6 light-days, respectively.}
    \label{ks_test_fdm_quad.pdf}
\end{figure*}

\begin{figure}
	\centering
	\begin{minipage}{\linewidth}
		\centering
		\includegraphics[width=\linewidth]{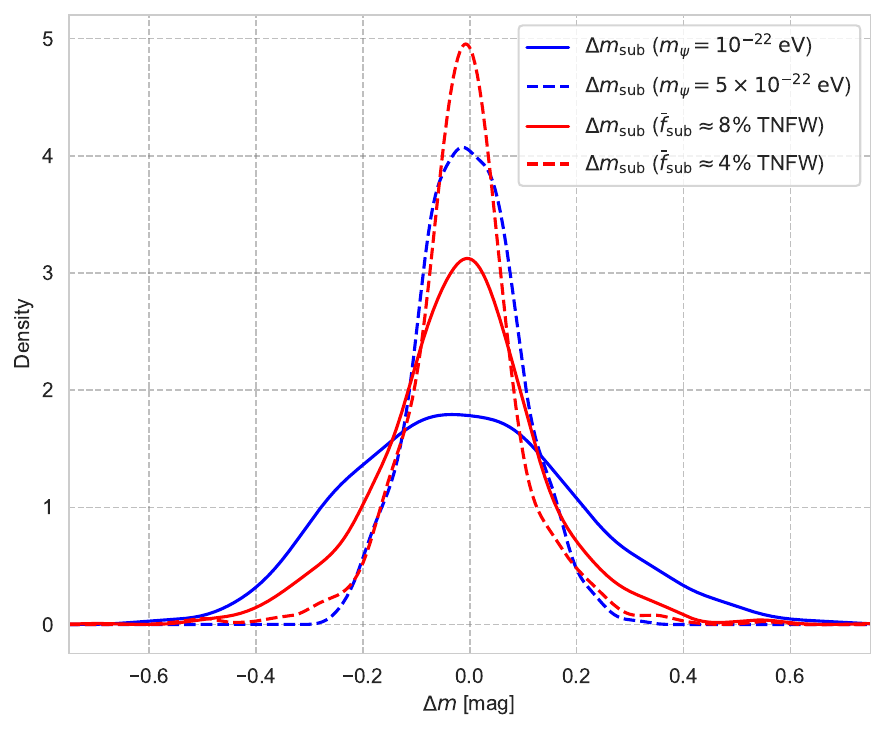}	
	\end{minipage}
    \caption{The probability density distributions of $\Delta m_\mathrm{sub}$, obtained using different types of dark matter substructure, for a strong lensing system with a dark matter halo of $M_{200} = 10^{12.35}\,\mathrm{M}_\odot$, concentration parameter $c_{200} = 9.5$, lens redshift $z_l = 0.789$, and source redshift $z_s = 1.74$. The solid and dashed blue curves correspond to FDM models with ultra-light boson masses $m_{\psi}=10^{-22}\,\mathrm{eV}$ and $m_{\psi}=5\times10^{-22}\,\mathrm{eV}$, respectively. The solid and dashed red curves represent CDM models with mean substructure mass fractions of $\bar{f}_{\rm sub}\approx8\%$ and $\bar{f}_{\rm sub}\approx4\%$, respectively, both assuming a TNFW profile. The plot highlights how different models produce distinct perturbation distributions.}
    \label{diff_sub.pdf}
\end{figure}

\section{RESULTS AND DISCUSSION}\label{4}
In Fig. \ref{ks_test_base_all.pdf} and Fig. \ref{ks_test_base_quad.pdf}, we use all strong lensing systems and only quads, respectively. 
In both figures, we set the null hypothesis $H_0$ such that the flux ratio perturbations arise solely from microlensing and observational uncertainties. 
The observed statistic $\Delta m$ is then drawn either from $H_0$ or from scenarios that include two fiducial dark matter substructure models (as described above), as well as CDM subhalos in the mass range $10^5 \le M_\mathrm{subhalo}/\mathrm{M}_\odot \le 10^{10}$ modeled with a TNFW profile. 
For realizations consistent with $H_0$, the resulting KS test p-values remains around $p_\mathrm{ks} \sim 0.5$ as the number of flux ratio observations increases, indicating consistency with the null hypothesis. 
As shown in Fig. \ref{ks_test_base_all.pdf} and Fig. \ref{ks_test_base_quad.pdf}, the results for the case with CDM subhalos in the range $10^7 \le M_\mathrm{subhalo}/\mathrm{M}_\odot \le 10^{10}$ are similar to those with subhalos in the range $10^5 \le M_\mathrm{subhalo}/\mathrm{M}_\odot \le 10^{10}$. 
This demonstrates that neglecting lower-mass subhalos does not affect our conclusions.

In Fig. \ref{ks_test_base_all.pdf}, when CDM subhalos are included, we find that the null hypothesis can be rejected at a significance level of 95 $\mathrm{per\,cent}$ when the number of flux ratio observations reaches approximately 1100 and 600 for accretion disk half-light radii of 3 and 6 light-days, respectively. 
For the FDM-induced fluctuation scenario with $m_{\psi} = 10^{-22}\,{\mathrm{eV}}$, the corresponding thresholds for rejecting $H_0$ are approximately 700 and 300 observations, respectively. 
These results indicate that as the number of optical flux ratio observations increases, perturbations from microlensing and observational uncertainties alone become insufficient to account for the observed flux ratio anomalies, necessitating the inclusion of dark matter substructure effects. 
The difference in the number of required observations for different accretion disk sizes arises because smaller disks are more strongly affected by microlensing \citep{2005ApJ...628..594M}. 
Consequently, distinguishing substructure-induced perturbations from microlensing-induced perturbations requires more observational data when the accretion disk is smaller. 

However, in Fig. \ref{ks_test_base_quad.pdf}, for the CDM subhalo scenario, only 90 and 50 flux ratio observations are required to reject the null hypothesis at the 95 $\mathrm{per\,cent}$ confidence level for accretion disk half-light radii of 3 and 6 light-days, respectively, which is significantly fewer than in the case using all strong lensing systems. 
For the FDM-induced fluctuation scenario with $m_{\psi} = 10^{-22}\,{\mathrm{eV}}$, the required number of flux ratio observations is also much smaller compared to the case using all strong lensing systems. 
The reason that quads alone require far fewer observations than all strong lensing systems is that doubles provide much weaker constraining power: on the one hand, substructure-induced perturbations are smaller for doubles, while on the other hand, double images may lie closer to the lens galaxy center, where microlensing-induced perturbations become stronger. 
Furthermore, we find that the ratio between the required number of flux ratios using all strong lensing systems and that using quads alone exceeds 5.5 (corresponding to 110 flux ratios for all systems versus 20 for quads). 
This demonstrates that including doubles not only fails to improve the constraints but even weakens them, implying that quads alone provide stronger constraints than using all strong lensing systems or using doubles alone (although we do not explicitly show the case of doubles only and the comparison between different dark matter substructure parameters in this paper, we have carefully verified them). 
In addition, we note that for the dark matter substructure parameters used in Figures \ref{ks_test_base_all.pdf} and \ref{ks_test_base_quad.pdf}, the perturbations induced by FDM on quads are similar to those from CDM subhalos, whereas for doubles the FDM-induced perturbations are stronger than those from CDM subhalos. We find that doubles have a higher fraction of lower-mass host halos compared to quads. Since FDM induces stronger perturbations in such low-mass halos, this leads to the more pronounced deviations observed in doubles.

In Fig. \ref{ks_test_subhalo_quad.pdf}, we only use quads and set the null hypothesis $H_0$ to correspond to a scenario in which CDM subhalos with a mean substructure mass fraction of $\bar{f}_\mathrm{sub} \approx 0.04$ are modeled using a TNFW profile. 
To demonstrate the potential of using optical flux ratios to distinguish different dark matter substructure parameters, the observed statistic $\Delta m$ is drawn either from $H_0$, or from two alternative models: (1) replacing the TNFW profile with the most compact form, a point mass profile; and (2) doubling the substructure mass fraction while keeping the TNFW profile unchanged. As in previous tests, when $\Delta m$ is drawn from the $H_0$ model, the KS test p-value remains close to 0.5, indicating no significant deviation from the null hypothesis.
In the case where the subhalos are modeled as point masses, we find that the null hypothesis can be rejected at the 95 $\mathrm{per\,cent}$ confidence level when the number of flux ratio observations reaches approximately 100 and 50 for accretion disk half-light radii of 3 and 6 light-days, respectively. 
For the case where the substructure mass fraction is doubled, the required number of observations increases significantly to about 1,100 and 500, respectively, to achieve the same level of statistical significance.
The reason fewer observations are needed to distinguish the point mass scenario is that extremely compact subhalos cause stronger perturbations in the flux ratios, making the deviation from the fiducial model more pronounced. 

In Fig. \ref{ks_test_fdm_quad.pdf}, we also use quads alone and set the null hypothesis $H_0$ to correspond to a scenario with FDM-induced fluctuations arising from an ultra-light boson with mass $m_{\psi} = 10^{-22}\,{\mathrm{eV}}$. 
The observed statistic $\Delta m$ is then drawn either from $H_0$, or from an alternative scenario in which the boson mass is increased to $m_{\psi} = 5 \times 10^{-22}\,{\mathrm{eV}}$.
As in previous analyses, when $\Delta m$ is drawn from $H_0$, the KS test p-value remains close to 0.5, indicating consistency with the null hypothesis. 
For the alternative scenario with the increased boson mass, the null hypothesis can be rejected at a significance level of 95 $\mathrm{per\,cent}$ when the number of flux ratio observations reaches approximately 400 and 180 for accretion disk half-light radii of 3 and 6 light-days, respectively.

Another interesting phenomenon shown in Fig. \ref{diff_sub.pdf} is that, for CDM subhalos, increasing the subhalo mass fraction does not significantly change the maximum perturbation, but primarily thickens the tail of the perturbation PDF, thereby enhancing the probability of large perturbations. 
In contrast, for FDM-induced fluctuations, an increase in fluctuation strength leads to both a larger maximum perturbation and a thicker PDF tail. 
This more dramatic change in the overall shape of the distribution makes flux ratio measurements provide stronger constraints on the FDM particle mass than on the CDM subhalo mass fraction.

\section{CONCLUSIONS}\label{5}
Due to the effects of stellar microlensing, only flux ratios measured in emission regions that are sufficiently extended to be unaffected by microlensing, such as the narrow-line region, radio emission from the background quasar, and the warm dust region of the quasar \citep{2019A&A...623A..79C,2020MNRAS.491.6077G,2024MNRAS.530.2960N,2024MNRAS.535.1652K}, can be reliably used to probe dark matter substructure.
To expand the available wavelength range for flux ratio studies and provide greater flexibility for future research, we confront the challenges posed by microlensing and propose a new method to detect dark matter substructure using optical flux ratios of strongly lensed quasars. 
We select 100 strong lensing systems consisting of 90 doubles and 10 quads to represent the overall distribution of strong lensing systems.
By comparing the distributions of optical flux ratios with and without dark matter substructure, or under different substructure models, and applying the Kolmogorov--Smirnov (KS) test, we demonstrate the feasibility of our proposed method.

We consider two representative types of dark matter substructure: CDM subhalos and FDM-induced fluctuations. 
Other forms of substructure, such as subhalos in warm dark matter (WDM) models, are also possible. 
While similar in nature to CDM subhalos, WDM substructure generally produces weaker perturbations due to the suppression of small-scale halo formation by free-streaming \citep{2016MNRAS.455..318B}.
Using quads alone provides better performance than using all strong lensing systems or only doubles. 
To distinguish between different dark matter substructure models, or to extract the perturbations induced by substructure from optical flux ratio measurements, several tens to a few hundred independent optical flux ratio measurements from quads are required. 
Each quad can provide two independent flux ratios per band. 
In addition, LSST will carry out its survey in six bands, and since the strength of microlensing depends on the observed wavelength, taking into account the six-band observations and the correlations between different bands, the required number of quads can be reduced by more than a factor of six \citep{2022ApJS..258....1B}. 
However, the exact number depends on various factors, such as the adopted observing strategy and the specific data analysis algorithms. 
Nevertheless, with the advent of next-generation wide-field surveys, the required sample sizes will be readily achievable in the near future \citep{2025OJAp....8E...8A}.

As an increasing number of strong gravitational lensing systems are being discovered, it is crucial to develop diverse methodologies to fully utilize the forthcoming large datasets. 
Our approach broadens the accessible wavelength range for studying dark matter substructure through flux ratio anomalies, thereby extending such studies beyond a few specific bands and, to some extent, relaxing the dependence on particular observational instruments.

\section*{Acknowledgements}
KL was supported by National Natural Science Foundation of China (No. 12222302) and National Key R\&D Program of China (No. 2024YFC2207400).
YG acknowledge the support from National Key R\&D Program of China grant Nos. 2022YFF0503404, 2020SKA0110402, and the CAS Project for Young Scientists in Basic Research (No. YSBR-092).

%%%%%%%%%%%%%%%%%%%%%%%%%%%%%%%%%%%%%%%%%%%%%%%%%%
\section*{Data Availability}
The data underlying this article will be shared on reasonable request to the corresponding author.

%%%%%%%%%%%%%%%%%%%% REFERENCES %%%%%%%%%%%%%%%%%%

\bibliographystyle{mnras}
\bibliography{ref}

%%%%%%%%%%%%%%%%%%%%%%%%%%%%%%%%%%%%%%%%%%%%%%%%%%

%%%%%%%%%%%%%%%%% APPENDICES %%%%%%%%%%%%%%%%%%%%%

%%%%%%%%%%%%%%%%%%%%%%%%%%%%%%%%%%%%%%%%%%%%%%%%%%

% Don't change these lines
\bsp	% typesetting comment
\label{lastpage}
\end{document}